# Visualizing Electronic Structure of Twisted Bilayer MoTe$_2$ in Devices

Cheng Chen[1*], William Holtzmann[2*], Xiao-Wei Zhang[3*], Eric Anderson[2], Shanmei He[1], Yuzhou Zhao[2,3], Chris Jozwiak[4], Aaron Bostwick[4], Eli Rotenberg[4], Kenji Watanabe[5], Takashi Taniguchi[6], Ting Cao,[3] Di Xiao[3,2], Xiaodong Xu[2,3], Yulin Chen[1]

[1]*Department of Physics, University of Oxford, Oxford, OX1 3PU, United Kingdom*
[2]*Department of Physics, University of Washington, Seattle, Washington 98195, USA*
[3]*Department of Materials Science and Engineering, University of Washington 98195, Seattle, Washington, USA*
[4]*Advanced Light Source, Lawrence Berkeley National Laboratory, Berkeley, California 94720, USA*
[5]*Research Center for Functional Materials, National Institute for Materials Science, Tsukuba, Japan*
[6]*International Center for Materials Nanoarchitectonics, National Institute for Materials Science, Tsukuba, Japan*
*These authors contributed equally to the work

The pursuit of emergent quantum phenomena lies at the forefront of modern condensed matter physics. A recent breakthrough in this arena is the discovery of the fractional quantum anomalous Hall effect (FQAHE) in twisted bilayer MoTe$_2$ (tbMoTe$_2$)[1-4], marking a paradigm shift and establishing a versatile platform for exploring the intricate interplay among topology, magnetism, and electron correlations. While significant progress has been made through both optical and electrical transport measurements, direct experimental insights into the electronic structure – crucial for understanding and modeling this system – have remained elusive. Here, using spatially and angle-resolved photoemission spectroscopy (μ-ARPES), we directly map the electronic band structure of tbMoTe$_2$. We identify the valence band maximum, whose partial filling underlies the FQAHE, at the K points, situated approximately 150 meV above the Γ valley. By fine-tuning the doping level via in-situ alkali metal deposition, we also resolve the conduction band minimum at the K point, providing direct evidence that tbMoTe$_2$ exhibits a direct band gap – distinct from all previously known moiré bilayer transition metal dichalcogenide systems. These results offer critical insights for theoretical modeling and advance our understanding of fractionalized excitations and correlated topological phases in this emergent quantum material.

The exploration of exotic quantum phenomena and emergent electronic states in two-dimensional (2D) materials has fueled a vibrant and rapidly advancing field in condensed matter physics[5]. Among these, twisted bilayer moiré systems have attracted intense interest as fertile platforms for uncovering novel electronic behavior[6-9]. Twist-induced moiré patterns can dramatically alter electronic properties, giving rise to remarkable phenomena such as unconventional superconductivity in magic-angle twisted bilayer graphene (MATBG)[10,11], and more recently, the observation of the fractional quantum anomalous Hall effect (FQAHE) in R-stacked twisted bilayer $MoTe_2$ (tb$MoTe_2$)[1-4]. The FQAHE – marked by the quantization of the anomalous Hall conductance at fractional values of the elementary charge – exemplifies a topologically nontrivial phase governed by strong electron correlations[12-15]. Understanding the underlying band structure of tb$MoTe_2$ is thus crucial for unraveling the mechanisms behind this exotic state and opens new avenues for both fundamental studies and potential technological applications.

In this study, we report a direct measurement of the electronic band structure of tb$MoTe_2$ with a 4° twist angle (the angle when FQAHE occurs) using angle-resolved photoemission spectroscopy with sub-micrometre spatial resolution (µ-ARPES). Over the past decade, µ-ARPES has emerged as a powerful technique for probing the electronic structures of 2D materials and their heterostructures[16-18]. However, most prior studies have focused on air-stable materials, such as graphene-based devices – including MATBG[19-22] – and stable transition metal dichalcogenides (TMDs)[23-28]. In contrast, devices based on air-sensitive materials like tb$MoTe_2$ are typically encapsulated with inert, stable layers such as graphene to prevent degradation. While graphene encapsulation is compatible with µ-ARPES by allowing photoelectrons to escape with ~30% efficiency through a monolayer graphene capping[29], and has been successfully applied to few-layer $WTe_2$[30,31], it remains incompatible with transport measurements and may introduce extrinsic effects such as strong dielectric screening and charge transfer, potentially obscuring the intrinsic electronic structures that give rise to the correlated topological states of the encapsulated material.

Instead, here we employ monolayer hexagonal boron nitride (hBN) to encapsulate the

tbMoTe$_2$ device. Owing to its wide band gap (~5.9 eV) and minimal thickness (~0.3 nm), monolayer hBN enables efficient photoelectron transmission in µ-ARPES measurements while preserving the intrinsic electronic structure of tbMoTe$_2$. A schematic of the µ-ARPES setup and device configuration is shown in Fig. 1a. The tbMoTe$_2$ device was fabricated entirely within an inert glovebox environment, with a partially overlapping graphene flake serving as an electrical ground (see Methods and Fig. S1 for fabrication details). The device is fully encapsulated between a monolayer hBN top layer and a ~12 nm thick hBN bottom layer (Fig. 1a).

Consistent with the optical image of the device (Fig. 1b(i)), the µ-ARPES spectral intensity integrated around the Fermi level ($E_F$) from real-space scanning (Fig. 1b(ii)) enables clear identification of regions with distinct conductivity, including gold contacts, the SiO$_2$/Si substrate, and hBN flakes. A higher-resolution scan of the sample area (Fig. 1b(iv); corresponding optical image in Fig. 1b(iii)) further employs the characteristic Te-$4d$ level photoemission signal to precisely locate the tbMoTe$_2$ region. The band structures of monolayer MoTe$_2$ and tbMoTe$_2$, shown in Figs. 2 and 3, were acquired at the spatial locations labeled as P1 and P2 in Fig. 1b(iv), respectively.

Remarkably, photoelectrons from the underlying MoTe$_2$ sample can also penetrate the monolayer hBN encapsulation layer efficiently, enabling high-quality ARPES measurements. Figure 2 displays well-resolved band dispersions for both monolayer MoTe$_2$ and tbMoTe$_2$. The 3D spectral plot (Fig. 2a) and constant energy contour (Fig. 2b) reveal the overall band structure within the first Brillouin zone (BZ), with the most prominent differences observed near the Γ point.

To highlight the distinctions between monolayer and twisted bilayer MoTe$_2$, detailed band dispersions along the M–Γ–K high-symmetry directions in the BZ are presented in Fig. 2c. At the Γ point, monolayer MoTe$_2$ exhibits a characteristic flat valence band top (Fig. 2c, top row). In contrast, tbMoTe$_2$ shows a pronounced upward shift of the valence band due to strong interlayer coupling, resulting in a dome-shaped valence band maximum (Fig. 2c, bottom row). At the K point, however, the band structures of both systems remain mostly similar, indicating small twist-induced modifications in that region of momentum space.

The different behavior at Γ and K points can be understood based on the orbital components of the bands at these two points (Fig. S2), respectively: In each constituent MoTe$_2$ layer, the K-valley bands are contributed by orbitals with in-plane spatial extensions (Mo $d_{xy}$, $d_{x2-y2}$ and smaller contributions from Te $p_x$, $p_y$ orbitals), which are weakly coupled between layers due to the small interlayer overlap of their wave functions; while for Γ valley, the bands are formed by Mo $d_{z2}$ and Te $p_z$ orbitals, which naturally have more out-of-plane extentions and larger overlap (thus stronger interlayer coupling), similar to the situation in twisted bilayer WSe$_2$ system[23].

The high-resolution measurements further enable the extraction of key band structure parameters. As shown in Fig. 2d, the valence band maximum (VBM) is located at the K point in both monolayer and tbMoTe$_2$ (see Fig. S3 for its robustness). Compared to monolayer MoTe$_2$ (Fig. 2d, top row), the valence band near the Γ point in tbMoTe$_2$ (Fig. 2d, bottom row) is significantly elevated due to interlayer coupling, bringing it closer in energy to the VBM at K. The identification of the VBM at the K point is the first step towards determining the magnitude and location of the band gap in tbMoTe$_2$.

To locate the conduction band minimum (CBM) of both MoTe$_2$ and tbMoTe$_2$, we need to probe the unoccupied bands above the Fermi level ($E_F$). This was achieved by employing in-situ electron doping via surface deposition of potassium, which effectively raises $E_F$ into the conduction band (Fig. 3a). Remarkably, this method is effective even in our monolayer hBN-encapsulated device. Using this approach, we observed the CBM emerging at the K point (Fig. 3b,c), unambiguarsly confirming the direct band gap nature of (4° twisted) tbMoTe$_2$ with an estimated direct band gap of ~1.1 eV – consistent with photoluminescence (PL) measurements under similar conditions (Fig. 3e and refs.[2,32]) – in stark contrast to all other known moiré systems based on transition metal dichalcogenides[33,34].

To gain a more comprehensive understanding of the electronic structure of tbMoTe$_2$, we performed first-principles calculations within the framework of Kohn-Sham density functional theory (DFT). The calculated band structures for monolayer MoTe$_2$, R-stacked bilayer MoTe$_2$, and tbMoTe$_2$ are presented and compared in Fig. 4. For monolayer MoTe$_2$, the calculations reveals a direct band gap at the K point of the Brillouin zone (Fig. 4a(i)), in excellent agreement with our experimental results (Figs. 2c and 3c, and Ref.[35]). For both R-stacked bilayer and tbMoTe$_2$ (Fig. 4a(ii, iii)), the valence band maximum at the Γ point is significantly elevated, consistent with the experimental data (Figs. 2c, 3c). However, the CBM in both bilayer configurations is predicted to shift away from the K point to an off-symmetry location (denoted as Q; Fig. 4a(ii, iii)), in clear contrast to our ARPES measurements, which reveal the CBM to reside at the K point (Fig. 3c).

To address this discrepancy[36,37], we investigated the effects of in-plane lattice parameters on the band structure. Our analysis reveals that the crystal structural parameters can significantly influence the electronic properties, and as a demonstration, we evaluate the effect of in-plane strain on the band structure. As illustrated in Figure 4b, the local conduction band minimum at the Q point shifts upward relative to the K point with a biaxial strain of 1%. This behavior can be understood by considering the orbital components contributing to the conduction band. At the K point, the conduction band states are dominated by $d_{z^2}$, whereas at the Q valley, they have a more mixed orbital character, including $d_{xy}$, $d_{x^2-y^2}$, and some chalcogen $p$ orbitals. As a result, the K valley is less sensitive to in-plane lattice expansion, but the Q valley shifts upwards more significantly. We found that with 1% of biaxial strain, we can align the CBM at the K point.

The successful extraction of the electronic band structure of tbMoTe$_2$ within a functional device with key band parameters derived from both our μ-ARPES measurements (Fig. S4) and DFT calculations (see Table 1), represents a significant step forward in elucidating the microscopic origins of the FQAHE. The fact that no apparent moiré or flat bands were observed near the VBM at the K point in current measurements suggests that the moiré potential in tbMoTe$_2$ is relatively subtle and highlights the robustness of the intrinsic band structure, which provides a valuable reference point for interpreting the emergent topological behavior. Looking ahead, we anticipate that future spectroscopic investigations – featuring higher energy and momentum resolution, improved statistical accuracy, and optimized device quality – will provide deeper insights into the complex interplay among lattice effects, topology, and strong correlations in moiré-engineered quantum materials.

In addition, this work demonstrates the feasibility of performing high-resolution μ-ARPES on hBN-encapsulated tbMoTe$_2$ devices, establishing a robust and non-invasive approach for probing the electronic structure of air-sensitive two-dimensional materials. Furthermore, the successful implementation of in-situ potassium dosing through the top hBN layer showcases a versatile and effective strategy for tuning the Fermi level and engineering the band structure *in operando*, particularly in situations where electrostatic gating is either impractical or insufficient.

## Methods:

### Sample fabrication

To fabricate the device, hBN and graphite flakes were first exfoliated onto SiO$_2$/Si substrates and characterized using contrast-enhanced optical microscopy and atomic force microscopy (AFM). Then the bottom gate structure was fabricated using a standard poly-(bisphenol A) carbonate (PC)-based dry transfer process. A hBN flake as the bottom gate dielectric was picked up, successively followed by a graphite bottom gate electrode and melted down onto a 285nm SiO$_2$/Si substrate with pre-patterned Cr/Au (7/70 nm) electrodes via standard electron beam lithography. Next, MoTe$_2$ flakes were exfoliated onto a 285nm SiO$_2$/Si inside a glovebox with O$_2$ and H$_2$O levels less than 0.1 ppm. A monolayer MoTe$_2$ flake was cut in half using an AFM tip before the transfer to minimize strain. The tbMoTe$_2$ heterostructure was created by picking up part of the monolayer MoTe$_2$ with the monolayer hBN encapsulation layer, rotating the transfer stage by a desired angle, and picking up the remaining flake. A graphene flake is also picked up, partially overlapped with tbMoTe$_2$ as a grounding electrode. The stack was then placed down on the prepared and cleaned backgate, such that the grounding graphene flake was electrically connected to the deposited contacts. The heterostructure geometry allows for full hBN encapsulation of the air-sensitive MoTe$_2$. The stamp polymer was dissolved in anhydrous chloroform for 5 minutes and washed by anhydrous isopropyl alcohol inside a glovebox. The sample surface was further cleaned by an AFM (Bruker Edge) in contact mode.

### Spatial- and Angle-resolved photoemission spectroscopy (μ-ARPES)

Synchrotron-based μ-ARPES measurements were performed at Beamline 7.0.2 (MAESTRO)

of the Advanced Light Source (ALS), USA. The samples were annealed in ultra-high vacuum at 220 °C for 3h and measured under ultra-high vacuum below $3\times10^{-11}$ Torr. The photon energy of the incident beam is 95eV (Fig. 2) and 70eV (Fig. 3), and the measurement was performed at a temperature of 20 K. Data was collected using a R4000 analyser upgraded with deflectors. The incoming photon beam was focused down to 2 um spot size by using a capillary mirror. The total energy and angle resolutions were 20 meV and 0.1°, respectively.

**Optical measurement**

Photoluminescence measurements were performed in reflection geometry in a home-built confocal optical microscope system. The sample was mounted in an exchange-gas cooled cryostat (attoDRY 2100) with a 9T superconducting magnet in Faraday geometry. A 632.8 nm HeNe laser was used as excitation for the photoluminescence measurements. The FWHM of the diffraction-limited excitation beam spot was ~1 μm. Photoluminescence signals were dispersed with a 600 grooves/mm diffraction grating blazed at 1μm and detected using a LN-cooled InGaAs photodiode array (Princeton Instruments PyLoN-IR 1.7). A long pass filter was used to remove the excitation laser from the photoluminescence signal before entrance into the spectrometer.

**Theoretical calculation**

The lattice relaxations of the moiré superlattice were performed using the neural network (NN) potential. The NN potential was parameterized using the deep potential molecular dynamics (DPMD) method[38,39]. The training datasets were generated from 5000-step *ab initio* molecular dynamics simulations (AIMD) at 500 K for a 6° tbMoTe$_2$ using the VASP package[40]. The van der Waals corrections were included using the D2 formalism[41]. More details of NN potential parameterization can be found in Ref. [[42]] . The experimental lattice constant of 3.492 Å was used for the unstrained monolayer, bilayer, and twisted bilayer MoTe$_2$[43]. The band structures were calculated within the SIESTA package[44]. Optimized norm-conserving Vanderbilt pseudopotentials[45], Perdew-Burke-Ernzerhof functional[46], and double-zeta plus polarization basis were used. The unfolding of moiré minibands into the extended Brillouin zone is performed following the method outlined in Ref. [[47]]. To simulate ARPES spectra, a Gaussian smearing is applied to the calculated band structure presented in Fig.4.

**Maintext Figures:**

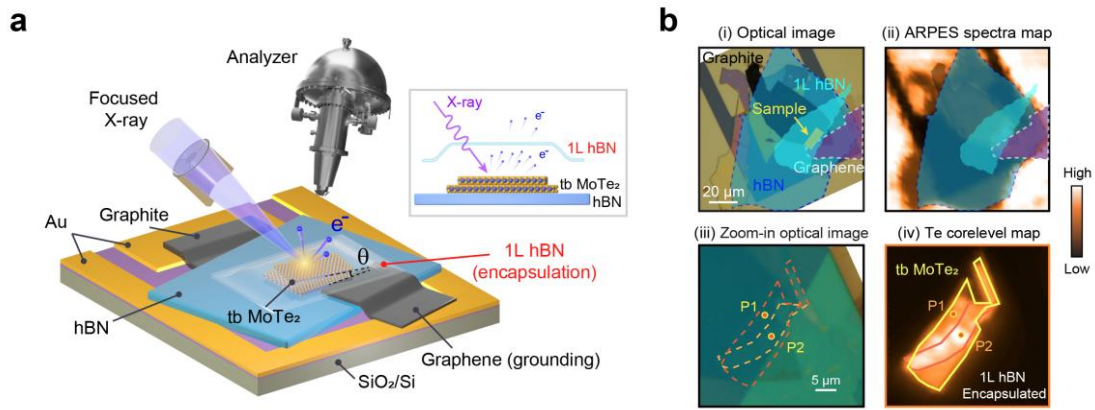

**Fig. 1| μ-ARPES measurement and tbMoTe$_2$ device geometries. a**, Schematic of the experiment setup. The tbMoTe$_2$ sample (4° twist) is fully encapsulated/protected by a monolayer hBN, and electronically grounded through a graphene/gold electrode. hBN/graphite backgate is also incorporated in this device (but not activated in the current study). **b**, Real-space information of the device. (i) Optical microscope image and (ii) corresponding ARPES spectral intensity map. The shape of encapsulation monolayer hBN, graphene electrode, and hBN/graphite backgate are indicated, respectively. (iii) Zoomed-in optical microscope image of the tbMoTe$_2$ device. Top/bottom MoTe$_2$ layers are highlighted with orange/yellow lines, and P1/ P2 indicates the spatial position of monolayer/twisted bilayer, where the μ-ARPES experiments were performed. (iv) Photoemission intensity map from Te-*4d* levels (E$_b$=40.4 and 41.9 eV for Te-*4d$_{5/2}$* and Te-*4d$_{3/2}$*, respectively), highlighting the tbMoTe$_2$ sample.

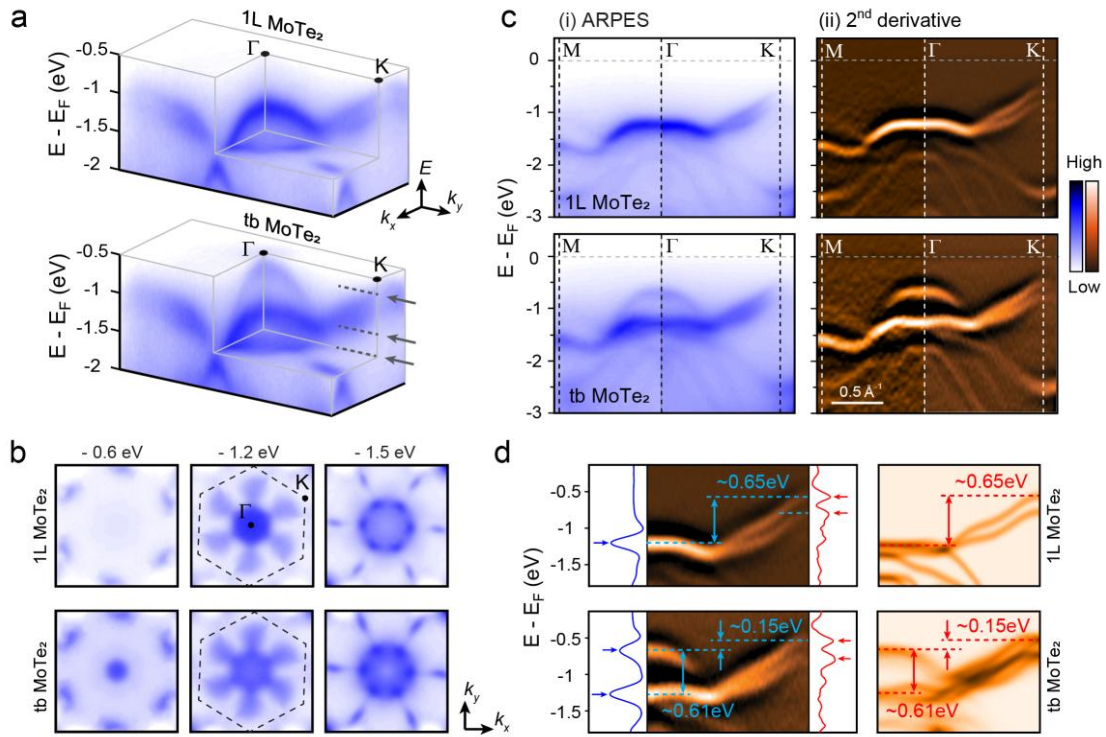

**Fig. 2| General electronic structure of monolayer and twisted bilayer MoTe$_2$. a,** Volume plot of photoemission intensity for monolayer (1L) MoTe$_2$ and twisted bilayer (tb) MoTe$_2$. High symmetry points Γ and K in the Brillouin Zone (BZ) are labelled. **b,** Constant energy contours of the photoemission maps at binding energies of -0.6 eV, -1.2 eV, and -1.5 eV, for 1L MoTe$_2$ and tb MoTe$_2$, respectively. The BZ and Γ and K points are overlaid. **c,** (i) ARPES band dispersion plot and (ii) corresponding second partial derivative plot, along M–Γ–K directions. **d,** Zoomed-in details of the band parameters (left column) and its comparison with theoretical calculation (right column). The intensity as a function of energy at both Γ (blue curve) and K (red curve) points are plotted on the side, with the band positions marked by the arrows.

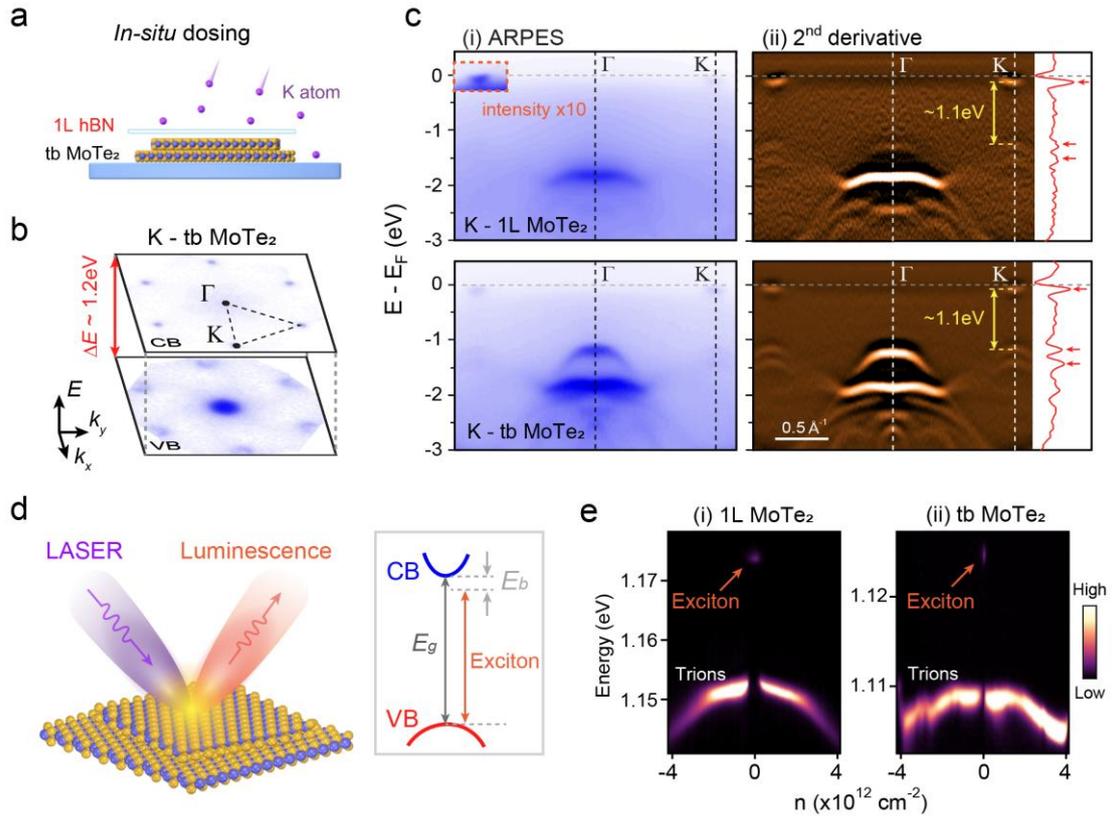

**Fig. 3| Direct band gap in monolayer and twisted bilayer MoTe₂. a,** Illustration of *in-situ* potassium surface dosing, which introduces additional electron carriers into the sample. **b,** Slices plot of constant energy contours of tbMoTe₂ after potassium surface dosing, showing the positions of conduction bands (CB) and valence bands (VB). **c,** (i) ARPES band dispersion plot and (ii) corresponding second partial derivative plot, along Γ–K direction of monolayer (top row) and tbMoTe₂ (bottom row), after potassium surface dosing. The intensity as a function of energy is plotted on the side in (ii), where the band positions are marked by arrows. **d,** Schematic of photon luminescence (PL) measurement and the generation of exciton. $E_g$: band gap size, $E_b$: exciton binding energy. **e,** PL intensity as a function of carrier density n, taken on (i) monolayer MoTe₂ and (ii) tbMoTe₂ devices (3.6° twist). See text for details.

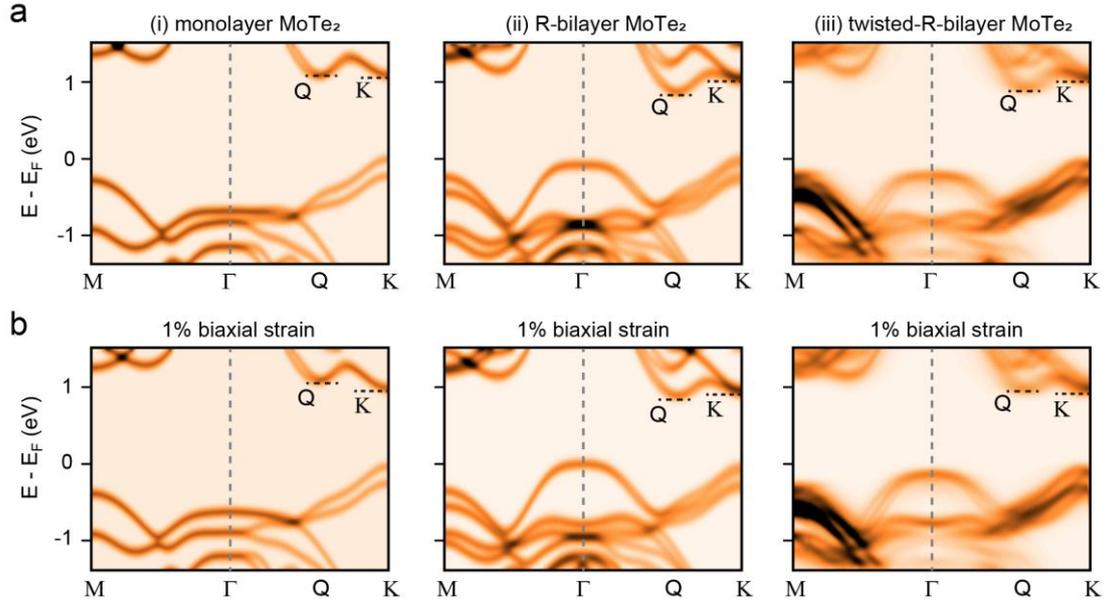

**Fig. 4| DFT calculated band structure. a,** Calculated band structure along M–Γ–K directions for (i) monolayer, (ii) R-bilayer, and (iii) twisted(3.89°)-R-bilayer MoTe$_2$. **b,** Calculated band structure with a biaxial strain of 1%.

**Main Text Table:**

|  |  | $E_g$ (eV) | $E_K - E_\Gamma$ (eV) | $\Delta_{hy}$ (eV) | $\Delta_{so}$ (eV) | $m_\Gamma^*/m_0$ | $m_K^*/m_0$ |
|---|---|---|---|---|---|---|---|
| Monolayer MoTe$_2$ | ARPES | 1.1 | 0.65 | - | 0.24 | 15.4 | 0.74 |
| | Calc. | 1.1 | 0.67 | - | 0.22 | 11.2 | 0.66 |
| 4° tbMoTe$_2$ | ARPES | 1.1 | 0.15 | 0.61 | 0.28 | 1.84 | 0.67 |
| | Calc. | 1.03 | 0.17 | 0.62 | 0.22 | 2.18 | - |

**Table 1|** Band parameters of monolayer MoTe$_2$ and tbMoTe$_2$, extracted from experimental results and theoretical calculations.